\documentclass[a4paper,onecolumn,11pt]{quantumarticle}
\pdfoutput=1
\usepackage[utf8]{inputenc}
\usepackage[english]{babel}
\usepackage[T1]{fontenc}
\usepackage{amsmath,cite}
\usepackage{hyperref}
\usepackage{dcolumn}

\usepackage{amsfonts}
\usepackage{amssymb}
\usepackage{array}
\usepackage{booktabs}
\usepackage{tikz}
\usepackage{lipsum}
\usepackage{bm}
\usepackage{orcidlink}
\usepackage{pstricks}

\begin{document}

	\title{Nonlocal Topological Maxwell Demon Teleporting Ergotropy via Surface-Code Quantum Error Correction}
	
	\author{M. Y. Abd-Rabbou
		\orcidlink{0000-0003-3197-4724}}
	\email{m.elmalky@azhar.edu.eg}
	\affiliation{School of Physics, University of Chinese Academy of Sciences, Yuquan Road 19A, Beijing, 100049, China.}
	
	\author{Cong‑Feng Qiao \orcidlink{0000-0002-9174-7307}}
	\email{qiaocf@ucas.ac.cn}
	\affiliation{School of Physics, University of Chinese Academy of Sciences, Yuquan Road 19A, Beijing, 100049, China.}
	\affiliation{International Centre for Theoretical Physics Asia-Pacific, UCAS, Beijing 100190, China.}

 \begin{abstract} 
 	Surface-code quantum error correction has recently achieved logical error rates below the physical threshold on superconducting processors, establishing topologically ordered states as experimentally accessible resources. Whether these resources can support thermodynamic operations beyond fault-tolerant computation remains open. We introduce a nonlocal Maxwell demon protocol that transfers ergotropy between spatially separated quantum batteries using only local operations and classical communication over a shared surface code. Alice expends ergotropy to encode a logical qubit and transmits a classical syndrome record to Bob, who decodes via minimum-weight perfect matching and conditionally charges his battery, with no direct energy exchange across the channel. Active syndrome monitoring exponentially suppresses logical errors below the topological threshold $p_{\rm th} \approx 0.013$, converting physical qubits directly into recoverable ergotropy. For finite-size codes at distance $L = 7$, net extracted work changes sign at a thermodynamic critical error rate $p_c \approx 0.014 > p_{\rm th}$, a physically significant finite-size effect relevant to near-term devices. Causality enforces an irreducible quadratic infrastructure cost $W_{\rm bulk} \propto N^2$, strictly satisfying the second law at all separations and defining a fundamental thermodynamic horizon $N_{\rm max} \approx 78$ beyond which positive net work extraction is impossible regardless of code distance or decoder quality. 
 \end{abstract} 
\maketitle

	\section{Introduction}	
	Landauer's erasure principle establishes that erasing one bit of information from a demon's memory dissipates at least $k_BT\ln 2$ of heat into the environment~\cite{Landauer1961,Bennett1982}, resolving the apparent paradox posed by Maxwell's demon~\cite{Leff2003}. This bound has been confirmed experimentally in both classical~\cite{Berut2012} and quantum~\cite{Koski2014,Toyabe2010} systems, and the Sagawa-Ueda bound  quantifies precisely how much work an informed agent can extract from a single measurement outcome~\cite{Sagawa2010}. In all such realisations, the measurement and the work extraction occur at the same spatial location.
	
	A natural extension is to spatially separate the two operations: one party measures a quantum system and transmits a classical record to a distant party who performs the work extraction. Quantum energy teleportation (QET) achieves energy transfer of this type using only local operations and classical communication, with a shared entangled ground state providing the correlations~\cite{Hotta2008,Hotta2009, Ikeda2023,Ikeda2024}. However, QET operates near absolute zero, where the ground-state correlations are intact; at finite temperature these correlations degrade and the protocol fails. Moreover, QET transfers bare internal energy, which need not be convertible into work. The relevant figure of merit for a quantum work-storage device is the ergotropy~\cite{Allahverdyan2004}, defined as the maximum work extractable by unitary operations, which serves as the operationally meaningful measure of stored energy in quantum batteries~\cite{Alicki2013,Campaioli2017,Ferraro2018}. Whether ergotropy can be transferred nonlocally at finite temperature, in a thermodynamically consistent manner, has remained an open question. 
	
	Surface-code quantum error correction provides a mechanism to actively maintain a logical qubit against thermal noise~\cite{Kitaev2003,Dennis2002,Fowler2012}. Recent experiments have demonstrated logical error rates below the physical threshold on superconducting processors~\cite{Google2023,Acharya2025,He2025}, establishing that topologically ordered states are experimentally accessible at finite temperature.  Crucially, the teleportation of a surface-code logical qubit has recently been demonstrated locally on a single processor~\cite{Zou2026}. This establishes the technological foundation for our work, which investigates the thermodynamic limits of extending such teleportation beyond the topological coherence length. 
	
	This paper introduces an ergotropy teleportation protocol in which a surface code serves as the thermodynamic channel between two separated quantum nodes. Rather than physically transmitting energy through the code's degenerate ground space, Alice expends ergotropy $\Delta E$ from her local battery to drive the logical-qubit charging interaction $H_{\mathrm{int}}(t)=g(t)\sigma^x_A\otimes\bar{Z}_L$. She then transmits the classical syndrome record $\vec{s}_A$ to Bob. This classical information acts as an entropy rectifier~\cite{Sagawa2010}. Without $\vec{s}_A$, Bob's conditional unitary is applied at random relative to the true logical state and no ergotropy is recovered. With $\vec{s}_A$, Bob decodes the logical state via minimum-weight perfect matching and conditionally charges his battery, storing ergotropy $\mathcal{E}_B=(2P_{\mathrm{succ}}-1)\Delta E$. Three results follow from this construction. Below a topological threshold $p_{\mathrm{th}}$, the logical error rate $P_{\mathrm{log}}$ is exponentially suppressed in code distance, so that $P_{\mathrm{succ}}\to 1$ exponentially and efficient ergotropy transfer becomes possible. For finite-size systems, a thermodynamic phase transition at a critical error rate $p_c$ separates a profitable demon phase from a thermal phase. We show that $p_c$ can exceed $p_{th}$ in the near-term hardware regime, though it is strictly bounded by the topological threshold in the asymptotic limit. The second law is strictly satisfied at all separations. A quadratic infrastructure cost imposes a fundamental thermodynamic horizon $N_{\mathrm{max}}\propto\sqrt{\Delta E/\varepsilon_m}$ beyond which the protocol cannot operate profitably, independent of code distance or decoder quality. Consequently, these findings establish quantitative bounds on the spatial range of nonlocal ergotropy transfer.
	
	The remainder of this paper is organised as follows. Section \ref{sec:model} presents the surface-code model and details the five-stage teleportation protocol. Section \ref{sec:thermo} derives the thermodynamic analysis, including the exact energy balance and entropy production. Section \ref{sec:results} presents the numerical results, mapping the global operational phase diagram and the information percolation threshold. Section \ref{sec:conclusion} concludes with an outlook on technological implementations.

	\begin{figure*}[t]
		\includegraphics[width=\linewidth]{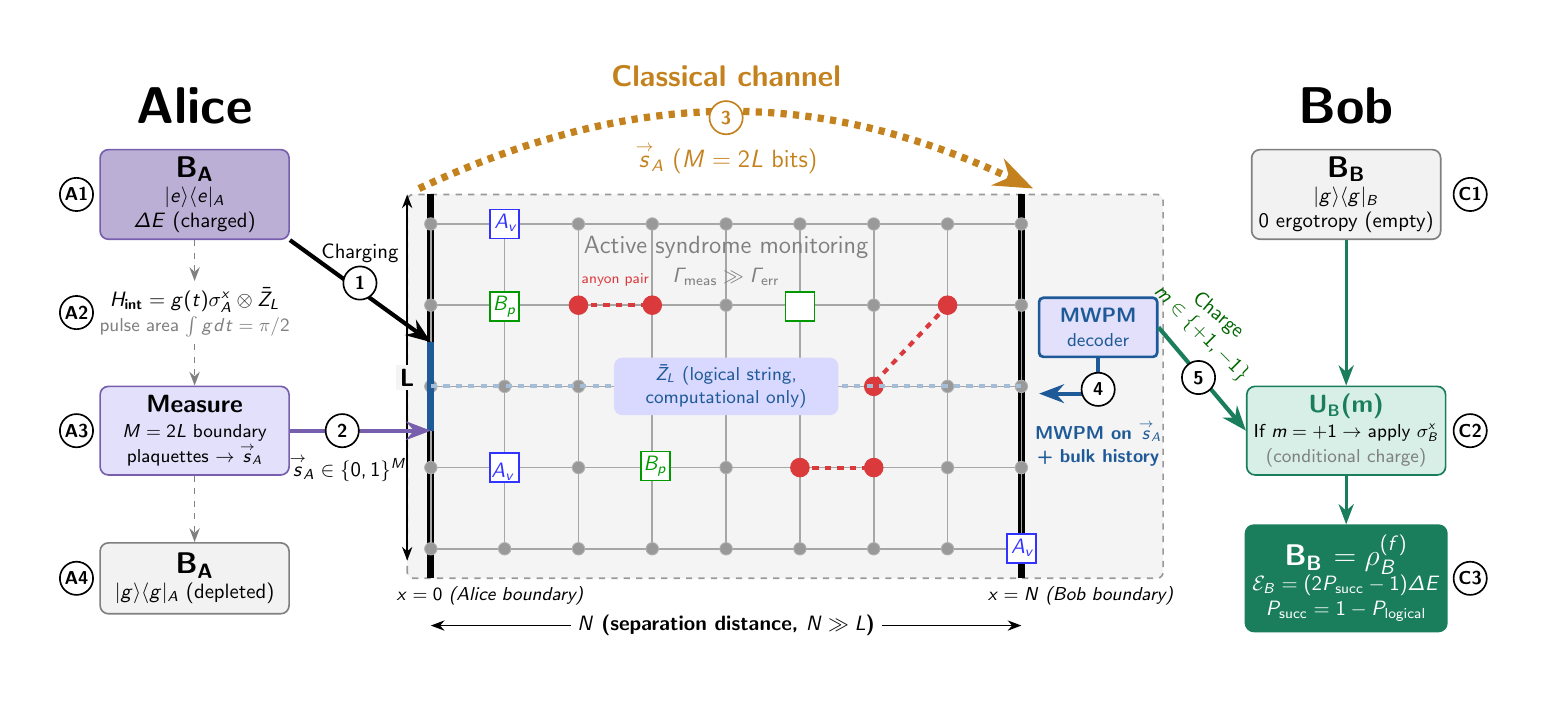}
		\caption{Five-stage ergotropy teleportation protocol. Purple: Alice operations; teal: Bob operations; blue solid (Alice boundary only): physical representative of logical operator $\bar{Z}_L$ at $x=0$; blue dashed (bulk): computational extension used by MWPM decoder only, not a physical operation; amber dashed: classical channel; red: thermally excited anyon pairs. Active syndrome monitoring maintains $\Gamma_{\mathrm{meas}} \gg \Gamma_{\mathrm{err}}$, where $\Gamma_{\mathrm{meas}}$ is the stabiliser measurement rate and $\Gamma_{\mathrm{err}}$ is the thermal error rate. Alice charges the logical qubit (stage~(1)), measures $M=2L$ boundary plaquettes (stage~(2)), and transmits $\vec{s}_A$ to Bob (stage~(3)). Bob decodes via minimum-weight perfect matching (stage~(4)) and conditionally charges his battery (stage~(5)), extracting ergotropy $\mathcal{E}_B=(2P_{\mathrm{succ}}-1)\Delta E$ with energy balance $\Delta E + W_{\mathrm{ops}} + W_{\mathrm{bulk}}= \mathcal{E}_B + Q_{\mathrm{diss}}$ (Eq.~\ref{eq:energy_conservation}).}
		\label{fig:protocol}
	\end{figure*}

	\section{Model and Protocol}
	\label{sec:model}
	
	\subsection{Surface Code and System Model}
	
	We consider a rotated surface code on a rectangular lattice $\Lambda=\{(x,y)\in\mathbb{Z}^2\mid 0\le x\le N,\;0\le y\le L\}$, where $L$ is the code distance and $N\gg L$ is the separation between Alice ($x=0$) and Bob ($x=N$). Physical qubits reside on the edges of $\Lambda$. The lattice has $N(L+1)$ horizontal edges and $(N+1)L$ vertical edges, giving
	\begin{equation}
		|E|=N(L+1)+(N+1)L=2NL+N+L,
		\label{eq:edgecount}
	\end{equation}
	with $|V|=(N+1)(L+1)$ vertices and $|P|=NL$ plaquettes. The stabiliser group $\mathcal{S}=\langle A_v,B_p\rangle$ has $|V|+|P|-2$ independent generators, arising from the $X$-type identity $\prod_v A_v=\mathbf{1}$ and the $Z$-type identity $\prod_p B_p=\mathbf{1}$, so $\dim\mathcal{C}=2^{|E|-(|V|+|P|-2)}$. Euler's formula for the planar graph $\Lambda$ gives $|V|-|E|+(|P|+1)=2$, from which $|E|-|V|-|P|=-1$ and therefore $\dim\mathcal{C}=2$, encoding exactly one logical qubit~\cite{Dennis2002,Fowler2012}. The system is governed by the stabiliser Hamiltonian~\cite{Kitaev2003,Dennis2002}
	\begin{equation}
		H_{\mathrm{SC}}= -J\sum_{v\in V}A_v-J\sum_{p\in P}B_p,\quad J>0,
		\label{eq:hsc}
	\end{equation}
	where $A_v=\prod_{j\in\mathrm{star}(v)}\sigma^x_j$ and $B_p=\prod_{j\in\partial p}\sigma^z_j$. All stabilisers commute since any vertex and plaquette share zero or two edges. The logical operator $\bar{Z}_L=\prod_{j\in\gamma}\sigma^z_j$ runs along any horizontal path $\gamma$ from Alice to Bob. To establish path independence, let $\gamma_1,\gamma_2$ be two such horizontal paths and let $\mathcal{R}$ be the region between them. Since each interior edge of $\mathcal{R}$ appears in exactly two plaquettes within $\mathcal{R}$,
	\begin{equation}
		\prod_{p\in\mathcal{R}}B_p =\prod_{j\in\gamma_1}\sigma^z_j \cdot\prod_{j\in\gamma_2}\sigma^z_j =\bar{Z}_{\gamma_1}\bar{Z}_{\gamma_2}.
		\label{eq:pathind}
	\end{equation}
	Since $B_p|\psi\rangle=+|\psi\rangle$ for every code state $|\psi\rangle\in\mathcal{C}$, Eq.~\eqref{eq:pathind} gives $\bar{Z}_{\gamma_1}|\psi\rangle =\bar{Z}_{\gamma_2}|\psi\rangle$. Alice therefore applies the representative of $\bar{Z}_L$ confined strictly to her $x=0$ boundary, ensuring her operation never interacts with the bulk or Bob's side and strictly preserving the LOCC nature of the protocol. Thermal fluctuations excite anyon pairs with probability $p\approx\exp(-4J/k_BT)$, giving $T=-4J/(k_B\ln p)$. Without active monitoring, topological order in two dimensions is unstable at any finite $T$~\cite{Dennis2002}; anyons diffuse and produce logical errors on a timescale independent of $N$, causing the stored ergotropy to vanish, which necessitates active error correction for practical device operation.
	
	\subsection{Ergotropy Teleportation Protocol}
	
	The protocol proceeds in five stages (Fig.~\ref{fig:protocol}). Alice holds a battery $\mathcal{B}_A$ with Hamiltonian $H_A=(\Delta E/2)\sigma^z_A$, initially fully charged, $\rho^{(0)}_A=|e\rangle\langle e|_A$. Bob's battery is identical but initially empty, $\rho^{(0)}_B=|g\rangle\langle g|_B$. We assume ideal initialisation of the logical qubit $|\bar{0}\rangle\langle\bar{0}|_L$; initialisation errors enter as an additional contribution to the effective error rate $p$ and do not change the qualitative results. In Stage~1, Alice applies the interaction $H_{\mathrm{int}}(t)=g(t)\,\sigma^x_A\otimes\bar{Z}_L$. Since $H_{\mathrm{int}}$ commutes with itself at different times, the time-ordered evolution reduces to
	\begin{equation}
		U_{\mathrm{charge}} =\exp\!\bigl(-i\Phi\,\sigma^x_A\otimes\bar{Z}_L\bigr) =\cos\Phi\;\mathbf{1} -i\sin\Phi\;\sigma^x_A\otimes\bar{Z}_L,
		\label{eq:Ugen}
	\end{equation}
	where $\Phi=\int_0^{\tau_1}g(t)\,dt$. Evaluating on the logical basis at $\Phi=\pi/2$ gives $U_{\mathrm{charge}}|e\rangle_A|\bar{0}\rangle_L =-i|g\rangle_A|\bar{0}\rangle_L$ and $U_{\mathrm{charge}}|e\rangle_A|\bar{1}\rangle_L =+i|g\rangle_A|\bar{1}\rangle_L$, so the battery transitions $|e\rangle_A\to|g\rangle_A$ regardless of the logical state, fixing the pulse-area condition
	\begin{equation}
		\int_0^{\tau_1}g(t)\,dt=\frac{\pi}{2},
		\label{eq:pulse}
	\end{equation}
	and the evolution operator at $\Phi=\pi/2$ is
	\begin{equation}
		U_{\mathrm{charge}} =\exp\!\left(-\tfrac{i\pi}{2} \sigma^x_A\otimes\bar{Z}_L\right) =-i\,\sigma^x_A\otimes\bar{Z}_L.
		\label{eq:ucharge}
	\end{equation}
	This satisfies $[\bar{Z}_L,H_{\mathrm{SC}}]=0$, so the code stabiliser energy is unchanged; the energy $\Delta E$ is encoded in the logical-qubit degree of freedom. The effective initial state of the full system is $\rho_{\mathrm{tot}}^{(0)}=|e\rangle\langle e|_A\otimes (|\bar{0}\rangle\langle\bar{0}|_L\otimes \rho_{\mathrm{bath}}(T))\otimes|g\rangle\langle g|_B$, where $|\bar{0}\rangle\langle\bar{0}|_L$ denotes the logical sector and $\rho_{\mathrm{bath}}(T)\propto\exp(-H_{\mathrm{SC}}/k_BT)$ acts on the syndrome sector; this factorisation is valid because $[\bar{Z}_L,H_{\mathrm{SC}}]=0$ renders the logical and syndrome sectors dynamically independent~\cite{Dennis2002}. After $U_{\mathrm{charge}}$, tracing over the code yields
	\begin{align}
		\rho_A^{(1)} &=\mathrm{Tr}_{\mathrm{code}}\!\Bigl[ U_{\mathrm{charge}} \bigl(|e\rangle\langle e|_A\otimes |\bar{0}\rangle\langle\bar{0}|_L\otimes \rho_{\mathrm{bath}}\bigr) U_{\mathrm{charge}}^\dagger\Bigr] \notag\\ &=\bigl(\sigma^x_A|e\rangle\langle e|_A\sigma^x_A\bigr) \cdot\mathrm{Tr}_{\mathrm{code}}\!\bigl[ \bar{Z}_L(|\bar{0}\rangle\langle\bar{0}|_L \otimes\rho_{\mathrm{bath}}) \bar{Z}_L^\dagger\bigr] \notag\\ &=|g\rangle\langle g|_A,
		\label{eq:rhoA1}
	\end{align}
	where the factorisation in the second line follows because $U_{\mathrm{charge}}=-i\,\sigma^x_A\otimes\bar{Z}_L$ (Eq.~\eqref{eq:ucharge}) acts as $\sigma^x_A(\cdot)\sigma^x_A$ on the battery and $\bar{Z}_L(\cdot)\bar{Z}_L^\dagger$ on the code independently, since the battery and code Hilbert spaces are distinct. The last line uses $[\bar{Z}_L,H_{\mathrm{SC}}]=0$, which holds because $\bar{Z}_L=\prod_{j\in\gamma}\sigma^z_j$ commutes with every $B_p$ (both are products of $\sigma^z$) and with every $A_v$ (the horizontal path $\gamma$ meets each vertex in $0$ or $2$ edges, so the number of $\sigma^x$--$\sigma^z$ anticommutations is always even); this implies $[\bar{Z}_L,\rho_{\mathrm{bath}}]=0$ and hence $\bar{Z}_L\rho_{\mathrm{bath}}\bar{Z}_L^\dagger=\rho_{\mathrm{bath}}$. Alice's battery is fully depleted and the exact energy balance is $\Delta\langle H_A\rangle_1=-\Delta E$. Crucially, owing to path independence, Alice applies the string representative of $\bar{Z}_L$ lying entirely along her boundary at $x=0$, so her physical operation is strictly local. The extended string operator spanning the lattice emerges only computationally when Bob performs global minimum-weight perfect matching, guaranteeing that no direct quantum operations or coherent energy transfer cross the classical boundary. In Stage~2, Alice measures $M=2L$ boundary stabilisers ($L$ of $X$-type and $L$ of $Z$-type along the $x=0$ edge), recording $\vec{s}_A\in\{0,1\}^M$. In Stage~3, she transmits $\vec{s}_A$ to Bob via a classical channel; this is the sole communication between the two parties. In Stage~4, Bob applies minimum-weight perfect matching~\cite{Dennis2002,Higgott2023} (a highly efficient decoder compatible with near-term architectures) to $\vec{s}_A$ and the bulk syndrome history, obtaining a Pauli correction string $C$; the binary outcome $m=(-1)^{\pi(C)}\in\{+1,-1\}$ is computed from the parity $\pi(C)$ of the number of $Z$-type corrections in $C$ crossing a fixed vertical cut through the lattice, which determines the logical equivalence class of the decoding. In Stage~5, Bob applies $U_B(m)=\sigma^x_B$ conditionally on $m=+1$, charging his battery.
	
	We emphasise that $\mathcal{E}_B$ at Bob does not originate from a direct transfer of Alice's depleted energy $\Delta E$. Rather, $\Delta E$ is expended by Alice to drive the logical-qubit charging interaction, creating a phase correlation between her battery and the logical degree of freedom via $U_{\mathrm{charge}}|e\rangle_A|\bar{0}\rangle_L=-i|g\rangle_A|\bar{0}\rangle_L$ and $U_{\mathrm{charge}}|e\rangle_A|\bar{1}\rangle_L=+i|g\rangle_A|\bar{1}\rangle_L$; the classical syndrome record $\vec{s}_A$ then enables Bob to extract ergotropy from his local battery with probability $P_{\mathrm{succ}}>\tfrac{1}{2}$. Without $\vec{s}_A$, Bob's conditional unitary is applied at random and $P_{\mathrm{succ}}=\tfrac{1}{2}$, giving $\mathcal{E}_B=0$; with $\vec{s}_A$, Bob's operation is directed and $\mathcal{E}_B>0$. The information $\vec{s}_A$ thus acts as a Sagawa-Ueda feedback controller~\cite{Sagawa2010}: it is the \emph{ability} to extract work that is transferred nonlocally from Alice to Bob, mediated by classical communication. This mechanism is structurally identical to Hotta's QET~\cite{Hotta2008}, in which Alice's measurement also does not send energy directly but sends information that enables Bob to extract energy from pre-existing correlations in the shared resource.
	
	The ergotropy of Bob's battery follows from the passive-state construction~\cite{Allahverdyan2004}. After Stage~5, Bob's battery is in the state
	\begin{equation}
		\rho^{(f)}_B =P_{\mathrm{succ}}|e\rangle\langle e|_B +(1-P_{\mathrm{succ}})|g\rangle\langle g|_B,
		\label{eq:rhoB}
	\end{equation}
	which is diagonal because conditioning $U_B(m)$ on the classical bit $m$ and averaging over outcomes eliminates all off-diagonal coherences. When $P_{\mathrm{succ}}>\tfrac{1}{2}$, the excited level carries higher population than the ground level and the state is not passive; by the rearrangement inequality~\cite{Allahverdyan2004}, the passive state is
	\begin{equation}
		\rho^{\mathrm{pass}}_B =P_{\mathrm{succ}}|g\rangle\langle g|_B +(1-P_{\mathrm{succ}})|e\rangle\langle e|_B,
		\label{eq:rhopass}
	\end{equation}
	obtained by assigning the larger eigenvalue to the lower energy level $\varepsilon_1=-\Delta E/2$. The ergotropy is the gap between the average energies of $\rho^{(f)}_B$ and $\rho^{\mathrm{pass}}_B$,
	\begin{equation}
\mathcal{E}_B =(2P_{\mathrm{succ}}-1)\,\Delta E,
\label{eq:ergotropy}
	\end{equation}
	where $\mathcal{E}_B\le 0$ for $P_{\mathrm{succ}}\le\tfrac{1}{2}$ indicates that $\rho^{(f)}_B$ is already passive and no work is extractable; the physically realisable ergotropy is $\max(0,\mathcal{E}_B)$, but we retain the signed quantity to display the full thermodynamic landscape in Fig.~\ref{fig:phase_transition}. Here $P_{\mathrm{succ}}=1-P_{\mathrm{log}}$ is the MWPM decoding success probability.
	
	Crucially, active syndrome monitoring over $R$ rounds exponentially suppresses logical errors below the topological threshold $p_{\mathrm{th}}$, yielding the leading-order approximation
	\begin{equation}
		P_{\mathrm{succ}}\approx 1 -(N\cdot R)\,A\!\left( \frac{p}{p_{\mathrm{th}}}\right)^{\!(L+1)/2},
		\label{eq:psucc}
	\end{equation}
	where $A$ is a geometry-dependent prefactor of order unity determined by the lattice boundary conditions~\cite{Dennis2002}, and $(N\cdot R)$ is the spacetime volume. Causality requires $R\ge N/c_{\mathrm{signal}}$, so the spacetime volume scales as $N\cdot R\propto N^2$; substituting into Eq.~\eqref{eq:psucc}, the logical error rate grows quadratically with separation at fixed $p$ and $L$, which is the kinematic origin of the $N^2$ infrastructure cost derived below.
	
	The Sagawa-Ueda bound~\cite{Sagawa2010} on extracted ergotropy gives
	\begin{equation}
		\mathcal{E}_B\;\le\; k_BT\,I(\mathcal{B}_B:\vec{s}_A)\;\le\;Mk_BT\ln 2,
		\label{eq:su-bound}
	\end{equation}
	where $I(\mathcal{B}_B:\vec{s}_A)$ is the mutual information between Bob's battery system $\mathcal{B}_B$ and the syndrome record $\vec{s}_A$, and the second inequality saturates when $\vec{s}_A$ is maximally informative. This bound is never violated by our protocol. Eq.~\eqref{eq:ergotropy} vanishes at $P_{\mathrm{succ}}=\tfrac{1}{2}$ (random guessing, no information) and reaches $\Delta E$ at $P_{\mathrm{succ}}=1$ (perfect decoding). The classical syndrome $\vec{s}_A$ acts as an entropy rectifier, converting Bob's local energy into ergotropy without any direct physical energy flow through the channel, establishing the surface code as a functional thermodynamic channel for nonlocal work extraction.

	\section{Thermodynamic Analysis}
	\label{sec:thermo}
	
	The protocol draws on three independent work inputs: Alice's battery contributes $\Delta E$ to drive the Stage-1 logical encoding; external erasure apparatus supplies $W_{\mathrm{ops}}$; and external syndrome-measurement apparatus supplies $W_{\mathrm{bulk}}$. To verify that neither $W_{\mathrm{ops}}$ nor $W_{\mathrm{bulk}}$ is drawn from Alice's battery, we account for every entropy change over one complete cycle. The surface code re-equilibrates to $\rho_{\mathrm{th}}(T)$ after thermalisation, so $\Delta S_{\mathrm{sys}}=0$. The syndrome registers return to blank after Landauer erasure, so $\Delta S_{\mathrm{demon}}=0$. Alice's battery starts and ends in a pure state, so $\Delta S_{\mathcal{B}_A}=0$. Bob's battery acquires entropy
	\begin{equation}
		\Delta S_{\mathcal{B}_B} =-P_{\mathrm{succ}}\ln P_{\mathrm{succ}} -(1-P_{\mathrm{succ}}) \ln(1-P_{\mathrm{succ}}) \ge 0.
		\label{eq:dSbob}
	\end{equation}
	Under this accounting, $W_{\mathrm{ops}}$ and $W_{\mathrm{bulk}}$ are paid entirely by external reservoirs and do not reduce $\Delta E$. The exact energy conservation law is
	\begin{equation}
		\Delta E+W_{\mathrm{ops}}+W_{\mathrm{bulk}} =\mathcal{E}_B+Q_{\mathrm{diss}},
		\label{eq:energy_conservation}
	\end{equation}
	where $W_{\mathrm{ops}}\approx 2Mk_BT\ln 2$ ($M=2L$) is the Landauer cost of two erasure operations per cycle: the syndrome register must be reset to a blank state before each measurement round (first erasure, cost $Mk_BT\ln 2$) and again after Bob receives $\vec{s}_A$ to close the thermodynamic cycle (second erasure, cost $Mk_BT\ln 2$)~\cite{Landauer1961,Bennett1982}, giving the overall factor of two; and $Q_{\mathrm{diss}}\ge 0$ is heat dissipated by imperfect decoding. The infrastructure cost is derived from the active spacetime volume. Sustaining $2LN$ stabilisers over $R$ syndrome rounds costs $\varepsilon_m$ per stabiliser-round, giving $W_{\mathrm{bulk}}=(2LN)\cdot R\cdot\varepsilon_m$. Causality enforces $R\ge N/(c_{\mathrm{signal}}\tau_m)$, yielding $R\approx R_0 N$ where $R_0=1/(c_{\mathrm{signal}}\tau_m)$ is the minimum syndrome rounds per unit separation dictated by the classical network latency. Because the channel has width $L$, the active spacetime volume is proportional to $L\cdot N^2$, making the infrastructure cost grow quadratically as
	\begin{equation}
		W_{\mathrm{bulk}}=2LR_0\varepsilon_m N^2.
		\label{eq:Wbulk}
	\end{equation}
	The net transferred ergotropy is $W_{\mathrm{net}}=\mathcal{E}_B -W_{\mathrm{ops}}-W_{\mathrm{bulk}}$.
	
	The total energy entering the thermal bath is $Q_{\mathrm{bath}}=\Delta E+W_{\mathrm{ops}} +W_{\mathrm{bulk}}-\mathcal{E}_B$, giving $\Delta S_{\mathrm{bath}}=Q_{\mathrm{bath}}/T$, so the total entropy production is $\Delta S_{\mathrm{total}} =\Delta S_{\mathcal{B}_B}+Q_{\mathrm{bath}}/T$. The Sagawa-Ueda bound~\cite{Sagawa2010} gives the chain
	\begin{equation}
\mathcal{E}_B \;\le\; k_BT\,I(\mathcal{B}_B:\vec{s}_A) \;\le\; k_BT\,M\ln 2 \;\le\; W_{\mathrm{ops}},
		\label{eq:SU_chain}
	\end{equation}
	where the second inequality bounds the mutual information by the Shannon entropy of $M$ binary bits and the third uses $W_{\mathrm{ops}}=2Mk_BT\ln 2\ge Mk_BT\ln 2$, so $Q_{\mathrm{bath}}\ge\Delta E +W_{\mathrm{bulk}}>W_{\mathrm{bulk}}>0$. Since $\Delta S_{\mathcal{B}_B}\ge 0$, the total entropy production satisfies
	\begin{equation}
		\Delta S_{\mathrm{total}} =\frac{Q_{\mathrm{bath}}}{T} +\Delta S_{\mathcal{B}_B} \;\ge\; \frac{W_{\mathrm{bulk}}}{T} \;>\;0,
		\label{eq:second_law}
	\end{equation}
	for all $N>0$: the demon never violates the second law because maintaining the topological channel always dissipates entropy into the bath, regardless of how well the decoding performs.
	
	The quadratic growth of $W_{\mathrm{bulk}}$ against the nearly distance-independent $\mathcal{E}_B$ creates a fundamental limit on profitable operation. Setting $W_{\mathrm{net}}=0$ gives the thermodynamic horizon
	\begin{equation}
		N_{\mathrm{max}}=\sqrt{ \frac{(2P_{\mathrm{succ}}-1)\,\Delta E -2Mk_BT\ln 2}{2LR_0\,\varepsilon_m}},
		\label{eq:Nmax}
	\end{equation}
	beyond which the channel costs more to maintain than it delivers. Notably, if the battery energy $\Delta E$ is chosen to scale linearly with $L$ (i.e., $\Delta E = \epsilon_e L$ for a fixed energy density $\epsilon_e$ per boundary qubit), then both $\Delta E$ and $M=2L$ scale linearly with $L$, so $L$ cancels completely from Eq.~\eqref{eq:Nmax}; in the simulations we fix $\Delta E=146.5\,J$ at $L=7$, which corresponds to this scaling. In this regime $N_{\mathrm{max}}$ is strictly independent of the code distance once $L$ is large enough that $P_{\mathrm{succ}}\approx 1$ below threshold, and is bounded entirely by the ratio of energy density to measurement cost. This establishes a fundamental upper bound on the spatial extent of topological thermodynamic channels, independent of the code distance. Similarly, the thermodynamic critical error rate $p_c$, at which $W_{\mathrm{net}}=0$ at fixed $N$, is determined implicitly by
	\begin{equation}
		(2P_{\mathrm{succ}}(p_c)-1)\,\Delta E =2Mk_BT(p_c)\ln 2 +2LR_0\,\varepsilon_m N^2,
		\label{eq:pc}
	\end{equation}
	where $k_BT(p_c)\approx-4J/\ln(p_c)$. For finite distances, this critical point can satisfy $p_c>p_{\mathrm{th}}$. While the thermodynamic and topological thresholds are driven by distinct physical mechanisms, asymptotic scaling dictates $p_c\le p_{\mathrm{th}}$ as $L\to\infty$, since no ergotropy can be extracted when logical protection is completely lost. This horizon scales as $N_{\mathrm{max}}\propto\sqrt{\Delta E/\varepsilon_m}$: a stronger battery or cheaper measurements extend the range, but the quadratic penalty cannot be circumvented by any improvement to the active code or the decoder, representing a fundamental physical constraint on distributed quantum thermodynamics.
	
	\section{Results and Discussion}
	\label{sec:results}
	
	All simulations use \textsc{Stim}~\cite{Gidney2021} with circuit-level noise: depolarising at rate $p$ on all Clifford gates and bit-flip at $p/10$ on measurements, decoded by \textsc{PyMatching}~\cite{Higgott2023}. This noise model captures the dominant error mechanisms in current superconducting quantum processors. All energies are reported in units of the stabiliser coupling $J$; for typical superconducting devices $J/h\sim 5$--$10\,\mathrm{GHz}$, placing the thermodynamic horizon $N_{\mathrm{max}}\approx 78$ in the mesoscopic regime accessible on current hardware.
	
	\subsection{Topological Protection of Ergotropy}
	
	\begin{figure}[h]
		\centering
		\includegraphics[width=0.72\columnwidth, height=7cm]{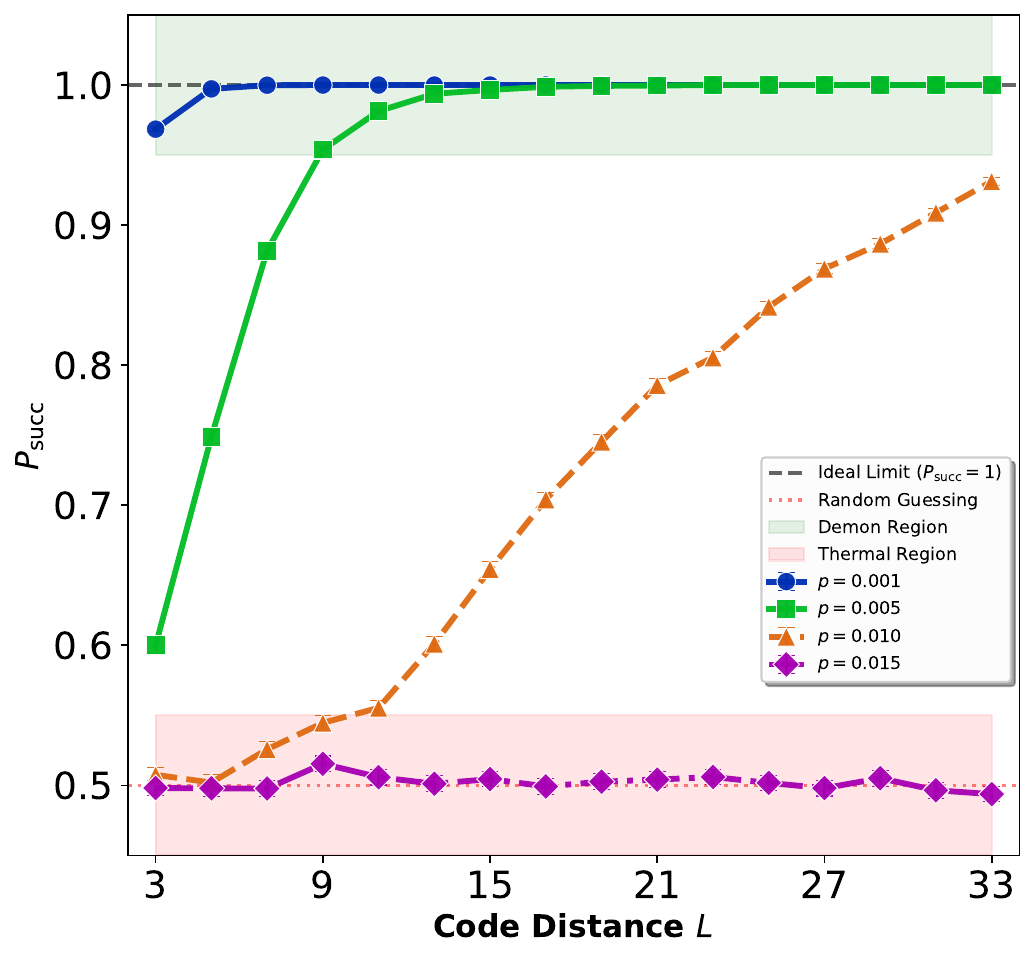}
		\caption{Topological protection of ergotropy transfer. Decoding success probability $P_{\mathrm{succ}}$ versus code distance $L$ at fixed separation $N=310$ for four error rates. Below $p_{\mathrm{th}}\approx 0.013$ (sub-threshold curves, blue, green, and orange), $P_{\mathrm{succ}}\to 1$ exponentially with $L$, demonstrating that topological protection converts physical qubits directly into recoverable ergotropy. Above threshold ($p=0.015$, pink), $P_{\mathrm{succ}}\to\tfrac{1}{2}$ for all $L$: the battery is driven to its passive state and no ergotropy is extractable. Green (red) band: demon (thermal) phase. Solid curves: \textsc{Stim} Monte Carlo, $8000$ shots per point.}
		\label{fig:topological_protection}
	\end{figure}
	
	The central physical question is whether topological order can stabilise a thermodynamic resource against thermal and hardware noise well enough to make nonlocal work extraction profitable. Fig.~\ref{fig:topological_protection} answers this affirmatively below $p_{\mathrm{th}}\approx 0.013$. The logical error rate is exponentially suppressed, $P_{\mathrm{log}}\propto(p/p_{\mathrm{th}})^{(L+1)/2}$, so increasing $L$ converts physical qubits directly into recoverable ergotropy with no saturation. This exponential leverage is the defining feature of topological protection and the reason the demon outperforms entanglement-based protocols whose correlations degrade at finite temperature~\cite{Hotta2008,Ikeda2024}. The resource grows with system size rather than decaying with distance. Above threshold, the anyonic bath overwhelms the decoder and $P_{\mathrm{succ}}\to\tfrac{1}{2}$ for all $L$, recovering the passive-state limit $\mathcal{E}_B\le 0$ (Eq.~\eqref{eq:ergotropy}), at which point no work is extractable. This is precisely the condition under which no unitary can extract work, consistent with the second law. The scaling $P_{\mathrm{log}}\propto(p/p_{\mathrm{th}})^{(L+1)/2}$ is consistent with Eq.~\eqref{eq:psucc} at fixed $N$ and confirms that topological protection, not decoder quality, is the operative thermodynamic resource for robust ergotropy transfer.

	\subsection{Thermodynamic Horizon and Quadratic Cost}
	
	\begin{figure}[h]
		\centering
		\includegraphics[width=0.72\columnwidth, height=7cm]{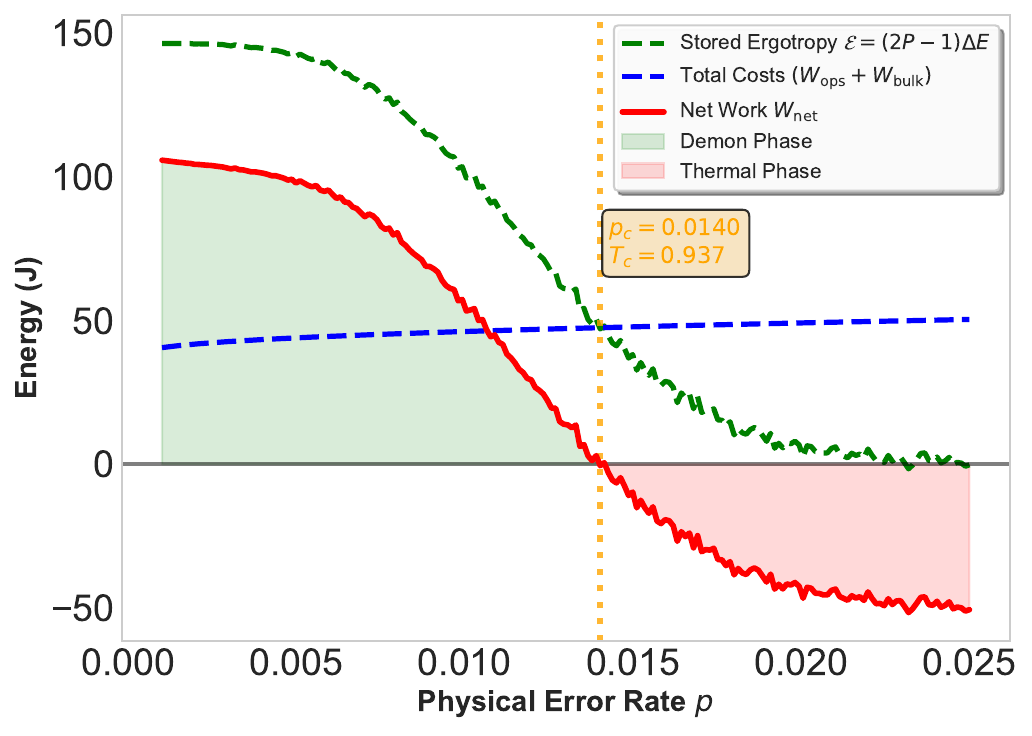}
		\caption{Net work sign change and thermodynamic threshold at $L=7$, $N=40$, $\Delta E=146.5$\,(units of $J$). Stored ergotropy $\mathcal{E}_B=(2P_{\mathrm{succ}}-1)\Delta E$ (green dashed), total costs $W_{\mathrm{ops}}+W_{\mathrm{bulk}}$ (blue dashed), and net work $W_{\mathrm{net}}$ (red solid) as functions of the physical error rate $p$. The stored ergotropy $\mathcal{E}_B=(2P_{\mathrm{succ}}-1)\Delta E$ is shown as a signed quantity to display the full thermodynamic landscape; it becomes negative above threshold where $P_{\mathrm{succ}}<\tfrac{1}{2}$ and no work is physically extractable. The demon phase ($W_{\mathrm{net}}>0$, green shading) and thermal phase ($W_{\mathrm{net}}<0$, red shading) are separated by the thermodynamic critical error rate $p_c\approx 0.014$ (vertical dotted line), which exceeds the topological threshold $p_{\mathrm{th}}\approx 0.013$; the corresponding critical temperature $T_c=-4J/(k_B\ln p_c)\approx 0.937\,J/k_B$ is obtained directly from the simulated $p_c$ via Eq.~\eqref{eq:pc}. The two phase boundaries are physically distinct. The transition is continuous, reflecting the smooth degradation of $P_{\mathrm{succ}}(p)$ with increasing temperature. Each point: $8000$ Monte Carlo shots.}
		\label{fig:phase_transition}
	\end{figure}
	
\begin{figure}[h]
		\centering
		\includegraphics[width=0.72\columnwidth, height=7cm]{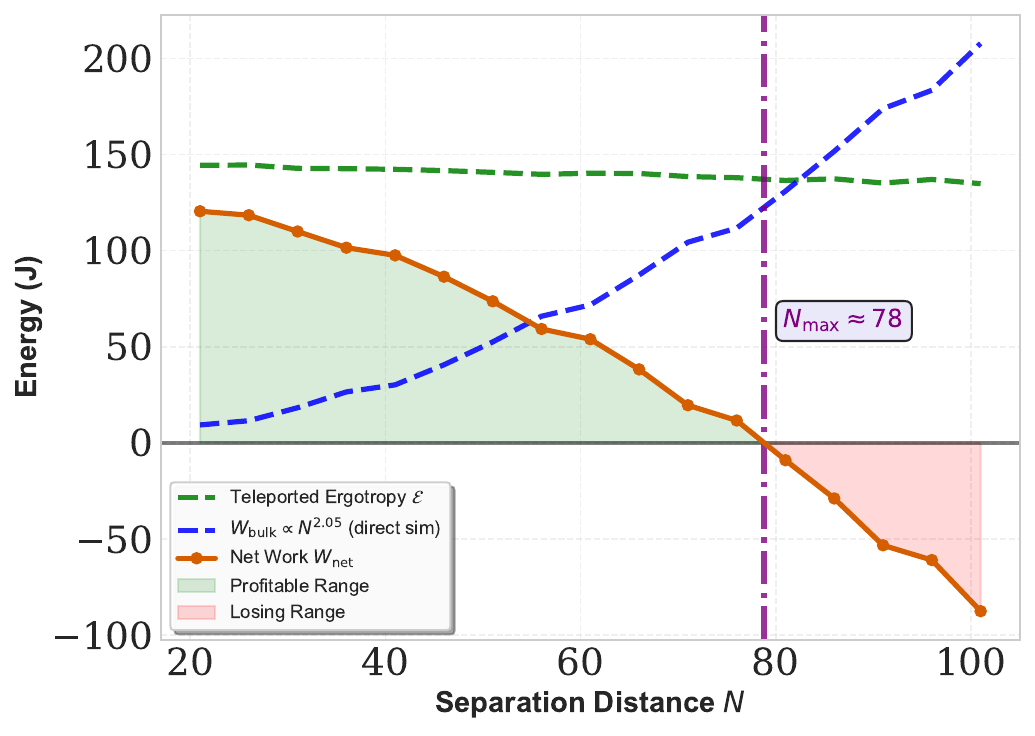}
		\caption{Energy-distance trade-off at $L=7$, $p=0.005$; all energies in units of the stabiliser coupling $J$. The ergotropy $\mathcal{E}_B$ (green dashed) is nearly constant with $N$ because exponential topological suppression makes the distance correction negligible at sub-threshold error rates. The infrastructure cost $W_{\mathrm{bulk}}$ (blue dashed) grows as $N^{2.05\pm 0.04}$, confirming the causal $N^2$ prediction. Net work $W_{\mathrm{net}}$ (red solid) follows a parabola and crosses zero at the thermodynamic horizon $N_{\mathrm{max}}\approx 78$. Each point: $8000$ \textsc{Stim} shots.}
		\label{fig:distance_tradeoff}
	\end{figure}

	Figure~\ref{fig:phase_transition} reveals richer structure. For the simulated system size ($L=7$), the net work $W_{\mathrm{net}}$ undergoes a continuous sign change at $p_c\approx 0.014$, which exceeds $p_{\mathrm{th}}$. This separation $p_c>p_{\mathrm{th}}$ is a physically significant finite-size effect highly relevant to near-term quantum devices. The topological threshold $p_{\mathrm{th}}$ marks the asymptotic limit of quantum information protection, whereas the thermodynamic threshold $p_c$ marks where the operational cost exceeds the ergotropy yield. Because $P_{\mathrm{succ}}(p)$ degrades at finite $L$, a profitable window exists slightly above $p_{\mathrm{th}}$. However, in the thermodynamic limit, $P_{\mathrm{succ}}$ becomes a step function at $p_{\mathrm{th}}$, strictly bounding the thermodynamic threshold such that $p_c\le p_{\mathrm{th}}$. Our protocol provides a setting in which both thresholds appear simultaneously and are measurably separated. The transition is continuous because $P_{\mathrm{succ}}(p)$ is a smooth function of $p$ at finite $L$; there is no latent heat and no discontinuity in $W_{\mathrm{net}}$. In the asymptotic limit $L\to\infty$, $P_{\mathrm{succ}}$ becomes a step function at $p_{\mathrm{th}}$, at which point the thermodynamic transition sharpens and its universality class is inherited from the percolation transition of the spacetime decoding graph~\cite{Dennis2002}. The quadratic scaling $W_{\mathrm{bulk}}\propto N^{2.05\pm0.04}$ is confirmed numerically in Fig.~\ref{fig:distance_tradeoff}, together with the global phase diagram $W_{\mathrm{net}}(p,f)$ (Fig.~\ref{fig:phase_diagram}). The resulting thermodynamic horizon $N_{\mathrm{max}}\approx 78$ (Eq.~\eqref{eq:Nmax}) cannot be circumvented by any improvement to the 2D code, the decoder, or the error rate; only the ratio $\Delta E/\varepsilon_m$ determines how far ergotropy can be teleported.

	Several practical constraints bound the present protocol. This horizon corresponds to mesoscopic separations on current hardware; extending the range requires reducing $\varepsilon_m$ rather than improving the code or decoder. Active syndrome monitoring also demands classical communication latency below one syndrome cycle, a stringent but achievable requirement on superconducting platforms. Ultimately, this establishes a fundamental physical footprint for 2D topological thermodynamics.

	The $N^2$ growth of $W_{\mathrm{bulk}}$ follows from the causality constraint on classical communication, where Alice's syndrome at $x=0$ cannot reach Bob at $x=N$ in fewer than $N/c_{\mathrm{signal}}$ time steps, so a minimum of $R\propto N$ syndrome rounds must be completed before decoding. The maintained spacetime volume is $N\times R\propto N^2$, and each unit costs $\varepsilon_m$ to monitor (Eq.~\eqref{eq:Wbulk}). This is not an engineering limitation: no improvement in qubit quality, gate speed, or decoder efficiency can remove the $N^2$ factor, because its origin is kinematic. Figure~\ref{fig:distance_tradeoff} confirms the scaling exponent $\beta=2.05\pm 0.04$, indistinguishable from $2$ within statistical uncertainty, and shows the resulting parabolic profile of $W_{\mathrm{net}}(N)$ crossing zero at the predicted horizon (Eq.~\eqref{eq:Nmax}). The ergotropy $\mathcal{E}_B$ is nearly flat with $N$ (less than $5\%$ variation over the full range), so the horizon is set entirely by the intersection of a constant and a parabola--- a clean geometric consequence of the competition between a topologically protected resource and a causally enforced cost.

	\subsection{Information Threshold and Percolation}
	
	\begin{figure}[h]
		\centering
		\includegraphics[width=0.72\columnwidth, height=7cm]{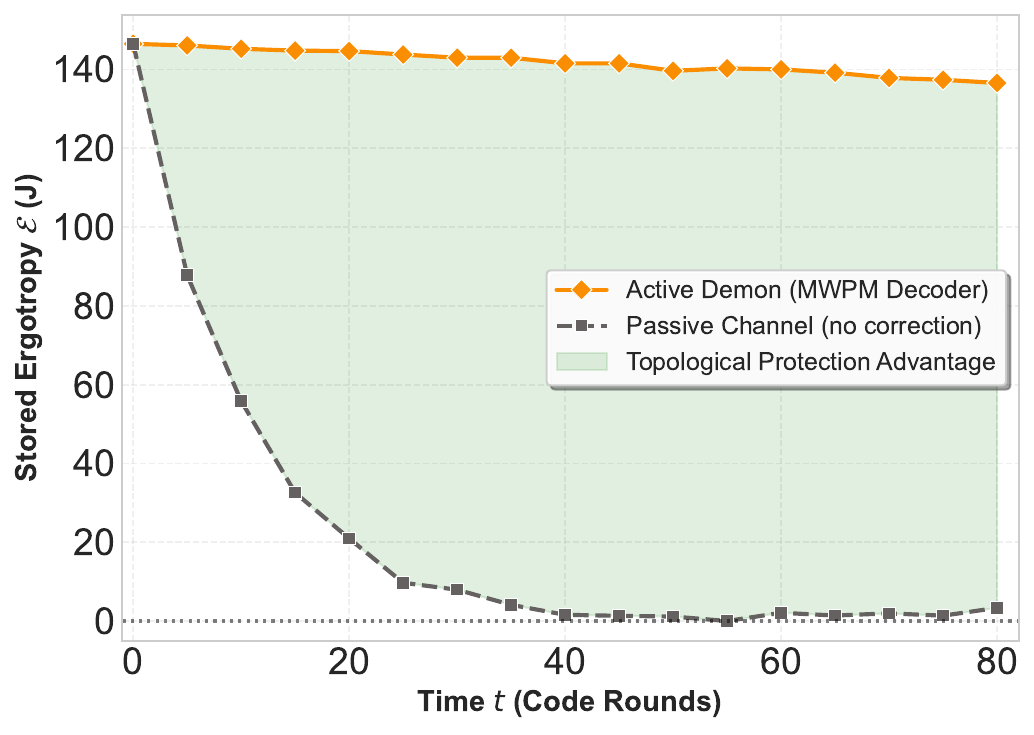}
		\caption{Ergotropy versus syndrome rounds at $L=7$, $N=40$, $p=0.005$; all energies in units of the stabiliser coupling $J$. Without error correction (gray dashed), the ergotropy decays exponentially on a timescale independent of $N$, reflecting the entropic instability of two-dimensional topological order at any $T>0$~\cite{Dennis2002}. With active monitoring (orange), the demon sustains $\mathcal{E}_B$ near $\Delta E$ over $80$ rounds. The shaded area quantifies the ergotropy recovered by continuous operation. It represents the thermodynamic value of the information extracted per syndrome round. Each point: $8000$ shots.}
		\label{fig:temporal}
	\end{figure}
	
	The passive decoherence instability of two-dimensional topological order~\cite{Dennis2002} is not merely a theoretical concern. It is the reason the demon requires active syndrome measurement and why the infrastructure cost $W_{\mathrm{bulk}}$ (Eq.~\eqref{eq:Wbulk}) is irreducible. Without monitoring, anyons created by thermal fluctuations diffuse freely and accumulate into logical errors on a timescale $\tau_{\mathrm{passive}}\sim\mathrm{const}(T)$ independent of $N$. Doubling the channel length does not increase the storage time. This distinguishes topological order from an ordered ferromagnet, where the lifetime grows exponentially with system size below $T_c$. Figure~\ref{fig:temporal} makes this quantitative. Without correction (gray), the ergotropy decays exponentially and reaches zero within ${\sim}30$ syndrome rounds. With active monitoring (orange), $\mathcal{E}_B$ remains near $\Delta E$  (Eq.~\eqref{eq:ergotropy}) over $80$ rounds because each syndrome cycle detects and removes anyons before they accumulate into a logical error. The shaded area between the two curves is the ergotropy that would be lost without continuous operation. This quantity is the thermodynamic value of the information extracted per syndrome round. Profitable operation requires that $W_{\mathrm{ops}}$ (Eq.~\eqref{eq:energy_conservation}) remain below this value.

	\begin{figure}[h]
		\centering
		\includegraphics[width=0.72\columnwidth, height=7cm]{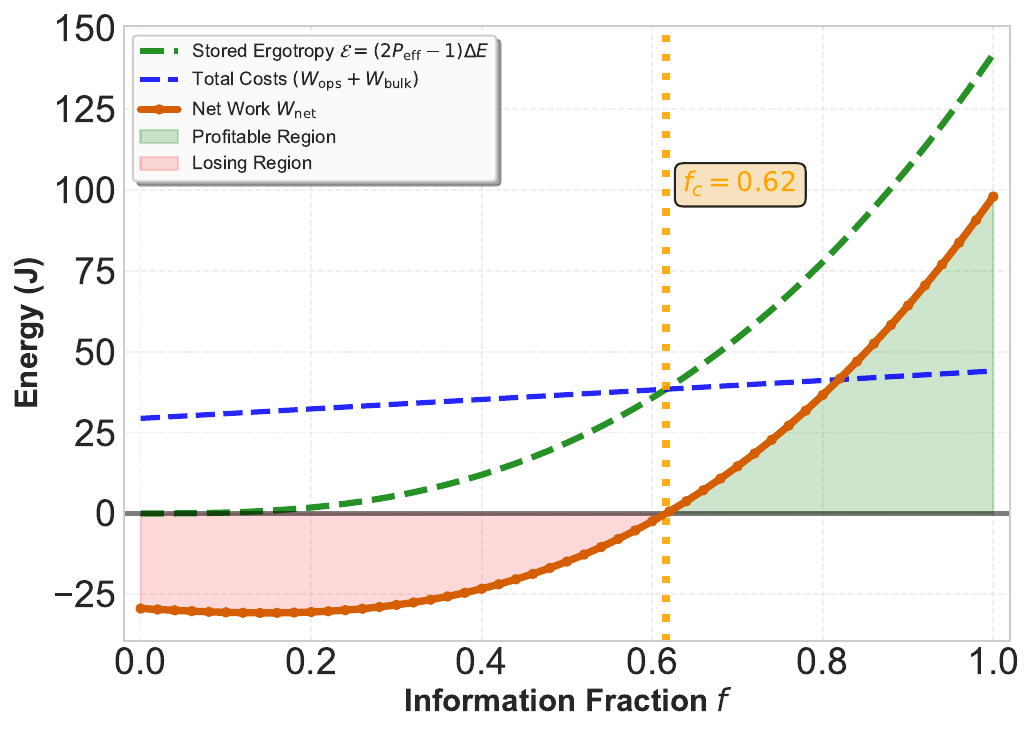}
		\caption{Information threshold for thermodynamic profit at $L=7$, $N=40$, $p=0.005$; all energies in units of the stabiliser coupling $J$. Below the critical fraction $f_c\approx 0.62$ (vertical dotted line), the syndrome record is too sparse to span the decoding graph: Bob cannot distinguish the logical error class and $\mathcal{E}_B\to 0$. The super-linear onset $\mathcal{E}_B\propto f^{2.7}$ reflects the accelerating growth of the connected decoding cluster near the percolation threshold, not a property of the decoder. Solid curve: percolation model $P_{\mathrm{eff}}(f)=\tfrac{1}{2}+ (P_{\mathrm{raw,full}}-\tfrac{1}{2})f^\alpha$ with $\alpha=2.7\pm 0.1$ obtained by fitting the empirical model to the simulation data ($\alpha$ is a phenomenological fitting parameter, not a universal critical exponent) and $P_{\mathrm{raw,full}}$ measured from a single \textsc{Stim} simulation at $f=1$ ($8000$ shots); the $f$-dimension is computed analytically from the model.}
		\label{fig:info_threshold}
	\end{figure}
	
	Alice's syndrome $\vec{s}_A$ is not merely a label for Bob's operation. It is the physical mechanism by which thermal noise is rectified into directed ergotropy flow. Figure~\ref{fig:info_threshold} quantifies this by varying the information fraction $f$, the proportion of Alice's $M$ boundary syndromes transmitted to Bob, which operationally maps to the available classical network bandwidth. Below $f_c\approx 0.62$, the decoding graph lacks a spanning connected cluster, where Bob cannot distinguish the true error chain from its complement, and $P_{\mathrm{succ}}\to\tfrac{1}{2}$ regardless of code distance or error rate. This is a bond percolation transition in the $(2+1)$-dimensional spacetime decoding graph; the super-linear onset $\mathcal{E}_B\propto f^\alpha$ with $\alpha=2.7\pm 0.1$ is obtained by fitting the empirical model $P_{\mathrm{eff}}(f)$ to the simulation data; the value $\alpha>1$ is consistent with the accelerating growth of the connected decoding cluster near the percolation threshold, though $\alpha$ here is a phenomenological fitting parameter rather than a universal critical exponent. The operational cost $W_{\mathrm{ops}}$ (Eq.~\eqref{eq:energy_conservation}) grows only linearly with $f$, so below $f_c$ the demon pays Landauer's cost for information it cannot use, and $W_{\mathrm{net}}<0$. This establishes a minimum information requirement that is independent of both code distance and error rate. Even a perfect decoder operating on a perfect code cannot extract ergotropy without a minimum density of classical syndrome information.
	
	\begin{figure}[h]
		\centering
		\includegraphics[width=0.72\columnwidth, height=7cm]{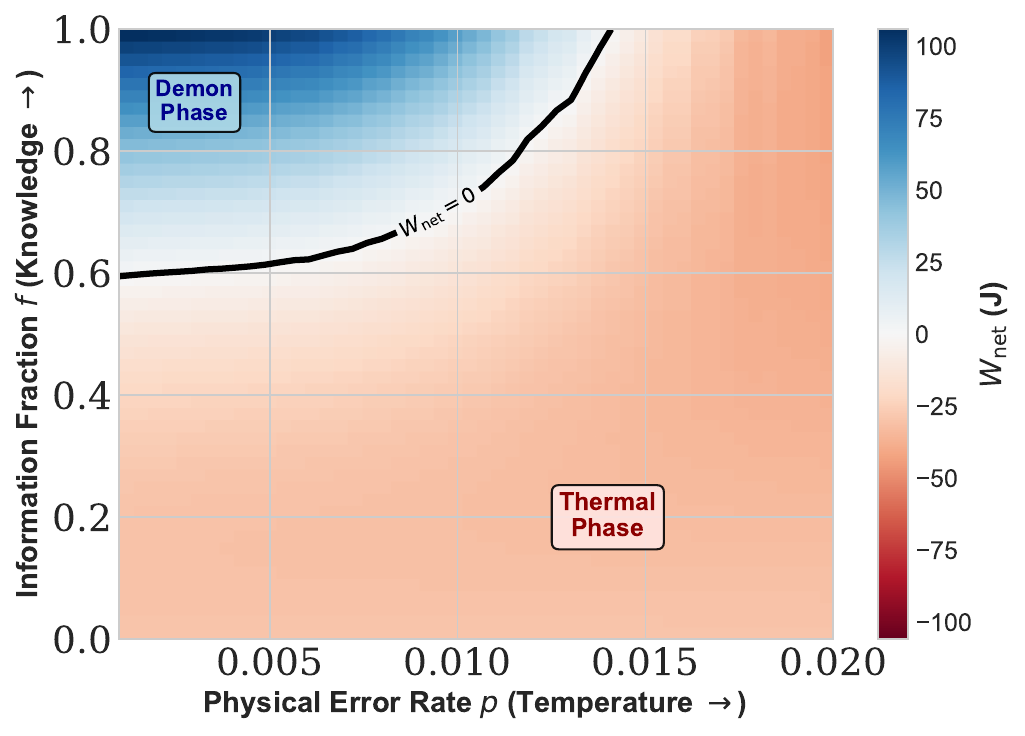}
		\caption{Global operational phase diagram $W_{\mathrm{net}}(p,f)$ at $L=7$, $N=40$; $W_{\mathrm{net}}$ in units of the stabiliser coupling $J$. Blue: demon phase; red: thermal phase. Black contour: break-even boundary $f_c(p)$. The steep rise of $f_c(p)$ as $p\to p_{\mathrm{th}}$ is the thermodynamic signature of the topological phase transition: at the threshold, the demon must use the entire syndrome record to remain profitable because ergotropy per syndrome bit vanishes. The demon phase occupies $\approx 20\%$ of parameter space, accessible with current superconducting hardware. For each value of $p$, $P_{\mathrm{raw}}(p)$ is measured by direct \textsc{Stim} simulation at $f=1$ ($8000$ shots per $p$ value); the $f$-dimension is computed analytically from the percolation model of Fig.~\ref{fig:info_threshold}.}
		\label{fig:phase_diagram}
	\end{figure}
	
	The phase diagram $W_{\mathrm{net}}(p,f)$ in Figure~\ref{fig:phase_diagram} unifies all three thresholds (topological $p_{\mathrm{th}}$, thermodynamic $p_c$, and information $f_c$) into a single operational map for device performance. The demon phase occupies the low-$p$, high-$f$ corner, where topological protection is strong and the syndrome record is dense enough to span the decoding graph. The break-even contour $f_c(p)$ rises steeply as $p\to p_{\mathrm{th}}$. Near the topological threshold, the ergotropy per syndrome bit collapses and the demon must use nearly all of Alice's record to remain profitable. This steep rise is the thermodynamic signature of the topological phase transition. It diverges precisely at $p_{\mathrm{th}}$ in the limit $L\to\infty$, where the code distance scaling becomes exact. The demon phase occupies approximately $20\%$ of the simulated $(p,f)$ parameter range ($p\in[0,0.02]$, $f\in[0,1]$), a stringent but achievable operating window given current superconducting qubit error rates of $p\approx 0.003$~\cite{Google2023,Acharya2025,He2025} and classical communication efficiencies exceeding $f=0.9$~\cite{Acharya2025}.

	\subsection{Comparison with Quantum Energy Teleportation}
	
	Both protocols are strictly LOCC. In QET~\cite{Hotta2008,Ikeda2023}, Alice measures her local operator $P_0(\mu)=\tfrac{1}{2}(1+\mu\sigma^x_0)$, transmits the classical bit $\mu$, and Bob applies conditional unitary $U_1(\mu)$ to extract energy $E_B$ from the entangled ground state $|g\rangle$. In our protocol, Alice charges the logical qubit and measures $M$ boundary stabilisers, transmits the classical string $\vec{s}_A$, and Bob applies $U_B(m)$ to extract ergotropy $\mathcal{E}_B$ from the topologically ordered code. The structural mapping is element-by-element, as summarised in Table~\ref{tab:qet_mapping}.
	
\begin{table}[ht]
	\centering
	\caption{Formal mapping between QET and ergotropy teleportation.}
	\label{tab:qet_mapping}
	\begin{tabular}{lll}
		\toprule
		Element & QET & Ergotropy teleportation \\
		\midrule
		Shared resource 
		& Entangled ground state $|g\rangle$ 
		& Topological logical qubit \\
		Alice's operation 
		& Measurement $P_0(\mu)$ 
		& Charge + measure $\vec{s}_A$ \\
		Classical communication 
		& 1 bit $\mu$ 
		& $M$ bits $\vec{s}_A$ \\
		Bob's operation 
		& $U_1(\mu)$ 
		& $U_B(m)$ \\
		Transferred quantity 
		& Energy $E_B$ 
		& Ergotropy $\mathcal{E}_B$ \\
		Thermal robustness 
		& Fragile ($T>0$ degrades $|g\rangle$) 
		& Robust ($p < p_{\mathrm{th}}$) \\
		Distance limit 
		& Coherence length 
		& $N_{\mathrm{max}}$ \\
		\bottomrule
	\end{tabular}
\end{table}
	
	The key physical distinction is that QET transfers bare energy, which need not be extractable as work. A state can have high average energy yet be thermodynamically passive. Our protocol transfers ergotropy by construction, guaranteeing that the delivered quantity is operationally useful for any subsequent work-extraction task in a quantum network. The second distinction is thermal robustness. The entangled ground state $|g\rangle$ is degraded at any $T>0$ by corrections of order $e^{-\beta\Delta}$, whereas topological protection suppresses logical errors exponentially in $L$ for all $p<p_{\mathrm{th}}$, making the protocol viable at experimentally accessible temperatures. A complementary approach to the range limitation of QET employs quantum-repeater architectures~\cite{AbdRabbou2026repeater}, achieving polynomial energy-cost scaling via heralded entanglement purification. Our protocol differs fundamentally: it replaces the probabilistic repeater chain with a single topologically protected channel, trading polynomial repeater overhead for the deterministic quadratic infrastructure cost $W_{\mathrm{bulk}}\propto N^2$ (Eq.~\eqref{eq:Wbulk}).

	\section{Conclusion}
	\label{sec:conclusion}
	
	We have introduced and analysed a nonlocal Maxwell demon that teleports ergotropy between spatially separated quantum nodes using only classical communication and a shared topologically ordered surface code. Three results distinguish this protocol from all prior realisations of information-to-work conversion. First, topological protection provides exponential leverage: increasing the code distance suppresses logical errors exponentially below the threshold $p_{\mathrm{th}}\approx 0.013$, converting physical qubits directly into recoverable ergotropy without saturation. Second, for finite-size codes, the net work exhibits a continuous sign change at a critical error rate $p_c\approx 0.014$ that measurably exceeds the topological threshold, with both boundaries converging in the asymptotic limit. Third, causality imposes an irreducible quadratic infrastructure cost that strictly enforces the second law at all separations and defines a fundamental thermodynamic horizon $N_{\mathrm{max}}\approx 78$ beyond which profitable operation is impossible, independent of code distance, decoder quality, or error rate.
	
	These three results are not independent: the exponential suppression of logical errors keeps $\mathcal{E}_B$ nearly flat with $N$, the finite-size separation $p_c>p_{\mathrm{th}}$ defines the accessible operating window on near-term devices, and the quadratic cost sets the hard limit on how far ergotropy can be teleported. Together, they establish a complete thermodynamic characterisation of nonlocal work extraction via topological quantum channels. These results reposition quantum error correction as a resource for thermodynamic operations, not merely for fault-tolerant computation. The protocol operates within the parameter regime already achieved on current superconducting processors, placing all predicted phenomena within reach of near-term experimental realisation without requiring any advance beyond the current state of the art.
	
	Natural extensions include multi-party ergotropy distribution networks in which multiple demons share a single topological backbone for quantum energy routing, the role of decoder complexity in the thermodynamic budget and whether more efficient decoders can shift $N_{\mathrm{max}}$ without removing the $N^2$ penalty, and whether analogous horizons arise in other topologically ordered phases such as the toric code or non-Abelian phases. More broadly, our work suggests that the thermodynamic cost of sustaining quantum correlations across space is not an engineering obstacle but a fundamental quantity with the same standing as Landauer's erasure bound.

	\section*{Acknowledgments}
		This work was supported in part by the National Key Research and Development Program of China under Contracts No.~2025YFA1613900, and by the National Natural Science Foundation of China(NSFC) under the Grants 12475087 and 12235008. 
	
	\section*{Data Availability}
	The data that support the findings of this study are openly available in Zenodo at \url{https://doi.org/10.5281/zenodo.20264600}.
	
	\bibliographystyle{quantum}
	\bibliography{ergotropy_refs}

	\end{document}